\documentstyle[11pt,amstex,righttag]{article}
\addtocounter{secnumdepth}{1}
\newcommand{\cN}{{\cal N}}
\newcommand{\cW}{{\cal W}}
\newcommand{\dd}{\text{d}}

\newcommand{\defin}{\equiv}
\newcommand{\CCP}{\text{\tiny CCP}}
\newcommand{\CP}{\text{\tiny CP}}
\newcommand{\R}{\text{\tiny R}}
\renewcommand{\text}{\mbox}

\begin{document}

\thispagestyle{empty}
\title{\bf Equivalence of stationary
state ensembles}
\vspace{10mm}
 
\author{H.J. Hilhorst and F. van Wijland\\
Laboratoire de Physique Th\'eorique$^1$\\
B\^atiment 210,
Universit\'e de Paris-Sud\\
91405 Orsay Cedex, France\\}

\maketitle

\begin{small}
\begin{abstract}
\noindent
We show that the contact process in an ensemble with conserved total
particle number, as simulated recently by Tom\'e and de Oliveira [{\it
Phys. Rev. Lett.} {\bf 86} (2001) 5463], is equivalent to the ordinary
contact process, in agreement with what the authors assumed and believed. 
Similar conserved ensembles and equivalence proofs
are easily constructed for other models.
\end{abstract}
\end{small}
\vspace{95mm}

\noindent LPT ORSAY 01/83\\
{\small$^1$Laboratoire associ\'e au Centre National de la
Recherche Scientifique - UMR 8627}
\newpage
%%%%%%%%%%%%%%%%%%%%%%%%%%%%%%%%%%%%%%%%%

A statistical mechanical ensemble is a probability distribution on a set
of microstates. 
For systems in thermodynamic equilibrium
the ensemble is known {\it a priori}. 
The equivalence of the standard equilibrium ensembles
(microcanonical, canonical, \ldots) is a well-known fact. 
In Monte Carlo simulations an easy way to generate a given ensemble is to
impose detailed balancing on the transition rates between the
microstates. 
A system of statistical
physics may be defined more generally, and more implicitly,
not by an {\it a priori} given
ensemble but by the transition rates
between the microstates. The stationary (= time
independent) solution of the resulting stochastic equation of motion
then constitutes an ensemble which, unlike the standard equilibrium
ensembles, cannot usually be expressed in explicit analytic
form. Such stationary ensembles \cite{note} arise, {\it e.g.}, in
reaction--diffusion type systems, of which one example is the object of
study here.\\

We consider the contact process
(CP) introduced by Harris \cite{Harris}, which takes place on an
infinite regular lattice each of whose sites may be empty or occupied by
a particle. The transition rates are such that particles (i) are created
on an empty site at a rate $n_{\rm nn}/z$, where $n_{\rm nn}$ is the
number of occupied nearest neighbors of that site and $z$ the lattice
coordination number; and (ii) annihilate spontaneously and
independently at a rate $k$. 

The contact
process is of interest because its stationary ensemble exhibits a
phase transition between an active state (particles present) and an
absorbing state (empty lattice) for a certain critical value $k=k_c$
of the annihilation rate \cite{note2}. 
Monte Carlo simulations of the active state on finite lattices and for
$k$ close to $k_c$ suffer from the complication that in any long enough
run there will inevitably occur a fluctuation
that pushes the system irreversibly
into the absorbing state. 
Various procedures have been proposed for dealing with this
``accidental death'', 
which occurs similarly in other model systems with
transitions to absorbing states.
Very recently Tom\'e and de Oliveira \cite{TomeOliveira} had recourse to
an elegant method,
used earlier in a different context by
Ziff and Brosilow 
\cite{ZiffBrosilow}, and which circumvents the accidental
death problem altogether.  Tom\'e and de Oliveira replace
the ordinary contact process (CP) by
a version with strictly conserved particle number
$N$. In this conserved contact process (CCP) particles (i) are created
at the same rate as in the CP; but (ii) every creation is accompanied by
the simultaneous annihilation of another particle picked at random on
the lattice.
It is our purpose here to 
show that the conserved ensemble (CCP) is equivalent to the
nonconserved ensemble (CP). This equivalence was assumed and believed to
be true by the authors \cite{TomeOliveira},
who nevertheless feel the need to invoke their results as an additional 
argument. The proof is simple and, just like the conserved
ensemble method itself, can easily be adapted to other problems.\\

Let $j$ denote a lattice site and let $n_j$ 
be its occupation number ($n_j=0$ or $=1$). It will be convenient to
work with ``spins'' $s_j=2n_j-1$, so that $s_j=\pm 1$. The
symbol $s\defin\{s_j\}$ then represents the set of occupation numbers.
We associate \cite{KadanoffSwift} with each $s$ a basis vector
$|s\rangle$ and with the time dependent
probability distribution $P(s,t)$ the {\it state vector}
\begin{equation}
|P\rangle_t=\sum_sP(s,t)|s\rangle
\label{statevector}
\end{equation}
The master equation for $P(s,t)$ is then equivalent to the evolution
equation
\begin{equation}
\frac{\dd |P\rangle_t}{\dd t}=\cW|P\rangle_t
\label{evolutioneq}
\end{equation}
for the state vector, in which the ``master operator'' $\cW$ is the
infinitesimal generator of the transitions. The $s_j=\pm 1$ may be
considered 
as the eigenvalues of a Pauli spin matrix $\sigma^z_j$. Since
$\sigma^x_j$ reverses $s_j$, one may express $\cW$ in terms of the
$\sigma^x_j$ and $\sigma^z_j$ \cite{Felderhof}. For the standard CP with
annihilation rate $k$ this yields $\cW=\cW^{\CP}$ with
\begin{eqnarray}
\cW^{\CP}&=&\cW_{\rm cre}\,+\,k\,\cW_{\rm ann}
\label{WCPWW}\\
\cW_{\rm cre}&=&\sum_j\sum_\delta{}'(\sigma^x_j-1)
\frac{1-\sigma^z_j}{2}\frac{1+\sigma^z_{j+\delta}}{2}\\
\cW_{\rm ann}&=&\sum_j(\sigma^x_j-1)\frac{1+\sigma^z_j}{2}
\label{WCP}
\end{eqnarray}
where $j+\delta$ is a neighbor site of $j$ and $\sum_\delta'$ denotes
$1/z$ times the sum on all neighboring sites. For the CCP with exactly
$N$ particles one finds $\cW=\cW^{{\CCP}}$ with
\begin{eqnarray}
\cW^{{\CCP}}=\frac{1}{N}\sum_j\sum_\delta{}'\sum_i
(\sigma^x_j\sigma^x_i-1)
\frac{1-\sigma^z_j}{2}\frac{1+\sigma^z_{j+\delta}}{2}
\frac{1+\sigma^z_i}{2}
\label{WCCP}
\end{eqnarray}
(in which the term $i=j$ is identically zero). The question is now to
demonstrate the equivalence, in a sense to be appropriately defined, of
$\cW^{\CP}$ and $\cW^{\CCP}$.
Our proof proceeds by two steps. The first one is to rewrite
expression (\ref{WCCP})
for the operator $\cW^{\CCP}$
by means of the substitution
\begin{equation}
\sigma^x_j\sigma^x_i-1=(\sigma^x_j-1)+(\sigma^x_i-1)
+(\sigma^x_j-1)(\sigma^x_i-1)
\label{substitution}
\end{equation}
Using that in the conserved ensemble $N=\sum_i(1+\sigma^z_i)/2$ we
find
\begin{equation}
\cW^{\CCP}\,=\,\cW_{\rm cre}+\cW_{\rm ann}\kappa +\cW_{\R}
\label{decomposition}
\end{equation}
Here $\kappa=\cN_{10}/N$ in which
\begin{equation}
\cN_{10}=\sum_j\sum_\delta{}'
\frac{1-\sigma^z_j}{2}\frac{1+\sigma^z_{j+\delta}}{2}
\label{defN10}
\end{equation}
is the operator for the total number of pairs of neighboring sites of
which one is occupied and the other empty; and the ``remainder''
$\cW_{\R}$ is given by
\begin{equation}
\cW_{\R}=\frac{1}{N}\sum_j\sum_\delta{}'\sum_i
(\sigma^x_j-1)(\sigma^x_i-1)
\frac{1-\sigma^z_j}{2}\frac{1+\sigma^z_{j+\delta}}{2}
\frac{1+\sigma^z_i}{2}
\label{defremainder}
\end{equation}
We remark that whereas $\cW_{\rm cre}$ and $\cW_{\rm ann}$ are
themselves master operators ({\it i.e.} correspond to master equations),
the remainder $\cW_{\R}$ is not; the reason is that
although it conserves the total probability, it does not conserve
the positivity of an initial probability distribution. 

The second step of the proof is to consider the time evolution of the
ensemble average of an arbitrary product 
$f\equiv\sigma^z_{\ell_1}\sigma^z_{\ell_2}\ldots\sigma^z_{\ell_r}$,
knowing that all physical observables are linear combinations of such
products. Let $\langle f\rangle^{\CCP}_t$ denote the average of $f$
in the conserved ensemble.
It may be calculated as the scalar product
\begin{equation}
\langle f\rangle^{\CCP}_t=\langle O|f|P\rangle_t
\label{faverage}
\end{equation}
where $|P\rangle_t=\exp(\cW^{\CCP}t)|P\rangle_0$ and where the
``projection state'' $\langle O|$ 
is defined by $\langle O|=\sum_s\langle s|$
with the sum running on all $s$ irrespective of the number of particles
present. From Eqs.\,(\ref{faverage}) and (\ref{decomposition})
it follows that
\begin{equation}
\frac{\dd\langle f\rangle^{\CCP}_t}{\dd t}\,=\,
\langle f\cW_{\rm cre}\rangle^{\CCP}_t
\,+\,\langle f\cW_{\rm ann}\kappa\rangle^{\CCP}_t
\,+\,\langle f\cW_{\R}\rangle^{\CCP}_t
\label{decaverageCCP}
\end{equation}
In the usual nonconserved ensemble one has in the same way from
Eqs.\,(\ref{faverage}) and (\ref{WCPWW}) that
\begin{equation}
\frac{\dd\langle f\rangle^{\CP}_t}{\dd t}\,=\,
\langle f\cW_{\rm cre}\rangle^{\CP}_t
\,+\,k\,\langle f\cW_{\rm ann}\rangle^{\CP}_t
\label{decaverageCP}
\end{equation}
We now compare Eqs.\,(\ref{decaverageCCP}) and (\ref{decaverageCP}).
In the limit of an infinite system 
(that is statistically invariant under translations)
the operator $\kappa$ will have
vanishingly small fluctuations around its average
$\langle\kappa\rangle^{\CCP}_t=
\langle\cN_{10}\rangle^{\CCP}_t/N$. Hence we may replace $\kappa$ by
this average. Next we will show that the last term on the RHS of
Eq.\,(\ref{decaverageCCP}) vanishes in the limit of an infinitely large
system. It is not sufficient for that to invoke the factor $1/N$ in
expression (\ref{defremainder}) for $\cW_{\R}$.
Instead, we will make the average
$\langle f\cW_{\R}\rangle^{\CCP}_t$ fully explicit.
In order to do so we commute the $\sigma^x_j$ through $f$ using that
$[\sigma^z_j,\sigma^x_j]_{_+}=0$, and use that
$\langle O|\sigma^x_j=\langle O|$. This gives
\begin{equation}
\langle f\cW_{\R}\rangle^{\CCP}_t\,=\,
\frac{4}{N}\,\sum_{p=1}^r\sum_{q=1}^r\sum_\delta{}'
\Big\langle f
\,\frac{1-\sigma^z_{\ell_p}}{2}\,\frac{1+\sigma^z_{{\ell_p}+\delta}}{2}
\,\frac{1+\sigma^z_{\ell_q}}{2}
\Big\rangle^{\CCP}_t
\label{fWRaverage}
\end{equation}
The RHS of this expression involves a sum on $r^2$ terms and the summand
makes no reference to the system size. Therefore it is now clear that in
view of the prefactor $1/N$ the average (\ref{fWRaverage}) 
vanishes in the limit
of infinite system size.
Hence Eq.\,(\ref{decaverageCCP})
may be replaced with
\begin{equation}
\frac{\dd\langle f\rangle^{\CCP}_t}{\dd t}\,=\,
\langle f\cW_{\rm cre}\rangle^{\CCP}_t
\,+\,\langle\kappa\rangle^{\CCP}_t
\,\langle f\cW_{\rm ann}\rangle^{\CCP}_t
\label{decaverageCCPbis}
\end{equation}
Comparison of (\ref{decaverageCP}) and (\ref{decaverageCCPbis})
shows that any spin average evolves according to the same equations of
motion in both ensembles provided we have 
$\langle\kappa\rangle^{\CCP}_t=k$, {\it i.e.}
\begin{equation}
\langle\cN_{10}\rangle^{\CCP}_t/N\,=\,k
\label{condition}
\end{equation}
This equation relates the CP parameter $k$ to a time dependent CCP
average. It can be satisfied only if that average is time independent,
hence equal to its stationary state value. Therefore the condition for
(\ref{decaverageCP}) and (\ref{decaverageCCP}) to be equivalent becomes
\begin{equation}
\langle\cN_{10}\rangle^{\CCP}_{\rm stat}/N\,=\,k
\label{statcondition}
\end{equation}
where the average is on the stationary CCP ensemble. In the
stationary state of the CCP, or for fluctuations around it that in the
large-$N$ limit affect $\langle\cN_{10}\rangle^{\CCP}_{\rm stat}/N$
negligibly,
equality (\ref{statcondition}) guarantees that the two averages 
$\langle f\rangle^{\CP}_t$ and $\langle f\rangle^{\CCP}_t$
obey the same equations.
It follows in particular that for any $f$ we have
$\langle f\rangle^{\CP}_{\rm stat}
=\langle f\rangle^{\CCP}_{\rm stat}$.
This establishes the equivalence of the two stationary ensembles. 
A notable case which on the basis of the preceding discussion is
excluded from the equivalence, is the relaxation towards
equilibrium of an
initial state in which all particles occupy random positions: In such a
nonequilibrium process \cite{note} the average of $\cN_{10}$
is not constant (in either ensemble)
and (\ref{statcondition}) cannot be satisfied.\\

The
equivalence condition (\ref{statcondition}) derived here was introduced
by Tom\'e and de Oliveira (Eq.\,(3) of \cite{TomeOliveira}), who exploit
it to determine the critical behavior of the contact process. 
Here we have furnished the proof (albeit a physicist's one) 
that their procedure was correct.

The construction of conserved ensembles by simultaneous execution of
elementary transitions at uncorrelated lattice locations 
can easily be adapted to other models of
interest in statistical physics. The same is true for the 
equivalence proof of this note.

%%%%%%%%%%%%%%%%%%%%%%%%%%%%%%%%%%%%%%%%%%%%%%%%%%%%%%%%%%%%%%%%%%%%%%%%%%%%%%


\begin{thebibliography}{10}
\bibitem{note}
The more general stationary
ensembles are referred to in Ref.\,\cite{TomeOliveira} by the term
``nonequilibrium''. 
Here we prefer to reserve that term
for time dependent ensembles.
\bibitem{note2}
The transistion is in the important ``Directed Percolation'' universality
class.
\end{thebibliography}
\end{document}